# Elephants, goldfishes and SOUL: a dissertation on forgetfulness and control systems

Guido Agapito*[a], Enrico Pinna[a], Alfio Puglisi[a], Fabio Rossi[a]

[a]INAF – Osservatorio Astrofisico di Arcetri, Largo E. Fermi 5, 50125, Firenze (FI)

## ABSTRACT

Adaptive Optics control systems accumulate differential measurements during closed loop operations to estimate turbulence and drive the deformable mirror. But have you ever wondered if your control system should be like an elephant, and never forget, or should it have a weak memory like a goldfish? Are measurement errors always zero mean or does static effects impact performance? Are commands high spatial frequencies good or are you wasting all the inter-actuator stroke for nothing? This work will try to answer these questions showing you results obtained during SOUL commissioning and analysing the impact of the values of the control system poles on Adaptive Optics. So be prepared to focus on forgetfulness and discover the advantages of being a goldfish in a digital world made of elephants.

**Note of the authors**:  and 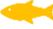 symbols are used in the text in place of "elephant" and "goldfish" words respectively.

## 1. INTRODUCTION

Memory is a key feature of humanity: human beings' ability to remember make them what they are, thus nowadays collective memory has emerged as one of the major topics, concerning the aim of linking the future generations to what the past have built. As Jan Assmann stated, "the specific character that a person derives from belonging to a distinct society and culture is not seen as a result of phylogenetic evolution, but rather as a result of socialization and customs" [1]. So many human creations deal with information that must be saved, stored and loaded. But even forgetting is important, as large scientific instruments select process to reduce the volume of data and collect only the essential data to be recorded (Compact Muon Solenoid (CMS) detector [2] which operates at the Large Hadron Collider (LHC) at CERN and Square Kilometer Array (SKA) [3] just to cite a few). Due the fact that "There is no firm psychological distinction between how we remember and how we think." [4], one may quest about how to manage a huge amount of data and if it is necessarily useful. One answer to this question comes from the act of forgetting, as it may have an adaptive value: when we have "to record and store all the stimuli we encounter, our memory would be a bedlam. So we choose, we filter" [5]. Following this perspective, forgetting therefore is the other coin of memory and both processes are do not differ too much: only a small amount of what we record can be recalled to our mind.

Adaptive Optics (AO) is no exception and AO closed loop control is a good symbol of the memorization process: the new closed loop control command is the sum of the past, the previous commands, and the current closed loop measurements (multiplied by appropriate gains).

Historically integrator has been the most used controller in Adaptive Optics: it is a simple first order low pass filter which gives good results in rejecting large and slow disturbances like turbulence. Both zonal or modal control have always relied on this kind of filter: easy to implement and with low computation requirement it is well suited for the high number of degrees of freedom of a typical astronomical adaptive optics system. Moreover, a single parameter for each filter must be optimized: the integrator gain.

SOUL [6] makes no exception: its real time computer has been designed to support a single pole filter, so little more than an integral control.

Integrator means memory, an infinite memory of the past… but our question is: is this infinite, infallible memory the best feature for our Adaptive Optics control systems? So, do we want an  control? Or is the weak memory of the 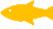 a desirable feature?

______________________________

*guido.agapito@inaf.it, Telephone: +39 0552752315

The paper is structured with a description of the disturbance of the AO loop in Sec. 2, a description of the control filter, results from SOUL commissioning in Sec. 4 and then some considerations on future and the conclusion in Sec. 5 and 6 respectively. Finally, we have two appendices in Sec. 7 presenting a method to estimate aliasing and calibration induced error and some example of the optimization of the control filter parameters.

## 2. DISTURBANCES

The AO control drives the Deformable Mirror (DM) to correct the wavefront distortion induced by atmospheric turbulence. The scheme of the closed loop is shown in Figure 1. Here we mean with disturbance those signals which disturb the slope measurements: they are the measurement noise, the spatial aliasing and the misregistration induced errors.

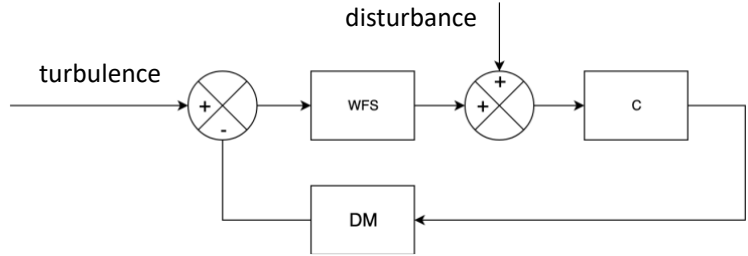

Figure 1. AO closed loop scheme. WFS is the Wave-Front Sensor, C is the Control, DM is the Deformable Mirror.

**Measurement noise** is the error induced by the photon propagation and the read-out of the detector. It is a white noise, that is a random signal having a constant power spectral density. As described in [7] the i$_{th}$ modal measurement noise, $\sigma_{w_i}^2$, for a Pyramid WFS, is:

$$\sigma_{w_i}^2 = p_i \sigma_s^2 \tag{1}$$

$$\sigma_s^2 = \frac{\sum_k x_k^2 \sigma_{I_k}^2}{(\sum_k I_k)^2} \tag{2}$$

$$\sigma_{I_k}^2 = F^2(n_k + b + d) + \sigma_r^2 \tag{3}$$

Where $p_i$ is the i$_{th}$ noise propagation coefficient from slope to mode, $x_k$ are the quad-cell weights, $I_k$ is the pixel intensity value, $F$ is the excess noise factor (to be accounted if the detector is an EMCCD), $n_k$ is the number of detected photons per frame of the k$_{th}$ pixel, $b$ is the number of detected photons per frame per pixel from the sky background, $d$ is the number of detected photons per frame per pixel from the dark current, $\sigma_r$ is the number of electrons RMS of read-out noise per frame per pixel.

**Aliasing error** is the spatial aliasing error made by the system WFS with a finite spatial sampling when it measures the residual wavefront incoming on the telescope pupil. So, spatial frequencies of the wavefront disturbances higher than the Nyquist frequency is sensed by the WFS as low spatial frequencies. Closed loop control is not able to distinguish this aliasing signal from the actual aberrations (for further reading on aliasing in Shack-Hartman WFS see [8]).

**Calibration induced errors** are those errors that originate from difference between the calibration set-up and the operation ones. For example, any misregistration of the DM actuators pattern with respect to the WFS sub-apertures pattern. These errors turn to a change in the WFS modal sensitivity generating signals that depend on different modes in a way similar to the spatial aliasing. In fact, we consider in the modal control [9] that each WFS modal measurement is orthogonal to the other modes, instead when a calibration error is present the orthogonality is no more real.

A method to estimate aliasing and calibration induced errors is presented in Sec. 7.

## 3. CONTROL FILTER

We present in this section the most common control filters used in AO, the integrator, and one of its variations, the leaky integrator, and we use the digital filter versions because AO systems are dealing with discrete-time signals. Note that the same considerations made for the integrator can be extended to any Infinite Impulse Response (IIR) filter.

The integrator (or pure integrator) can be expressed as:
$$y(k+1) = y(k) + gu(k+1) \quad (4)$$
$$C(z) = \frac{g}{1-z^{-1}} \quad (5)$$
where $u$ is the input signal, $y$ is the output signal, $k$ is the time step, $g$ is a free parameter and $z$ is the variable of the Z-transform. The only parameter of this filter is the gain $g$. This kind of integrator is the 🐘 integrator that never forgets the old integrated values $y$. The effect of changing the gain on the closed loop Transfer Functions (TF) are shown in Figure 2.

The leaky integrator adds a new parameter, the forgetting factor $f$ ($f < 1$):
$$y(k+1) = fy(k) + gu(k+1) \quad (6)$$
$$C(z) = \frac{g}{1-fz^{-1}} \quad (7)$$
The name of this parameter comes from the fact that $1-f$ of the old integrated values (output $y$) is "forgotten" at each iteration. This kind of integrator is the 🐟 integrator that apply $f^N$ coefficient to integrated values of $N-1$ steps old: even $f$ values slightly smaller that 1 makes negligible the contribution of integrated values old few hundreds of steps. The effect of $f$ on the closed loop Transfer Functions (TF) are shown in Figure 3. Note that the leaky integrator is used in several AO systems and few of the many available references in our field are [10],[11][12],[13] and [14].

So, $g$ has a direct connection with bandwidth and noise propagation on high temporal frequencies and it is effective in reducing noise propagation, but at the cost of bandwidth. Instead $f$ has a direct connection with steady state rejection and noise propagation on low temporal frequencies and it increases the bandwidth. Moreover, while phase and gain margins decrease when $g$ increases, smaller $f$ values correspond to greater stability: for example, the TF reported in Figure 3 with $g = 0.5$ and $f = 0.9$ has a phase margin of 44deg, 10deg more than the pure integrator.

These characteristics makes the pure integrator well suited for low order modes that has high turbulence power and high ratio turbulence/aliasing, while leaky integrator is well suited for high order modes that has low turbulence power and low ratio turbulence/aliasing. This conclusion is confirmed by the examples of integrator gain and forgetting factor optimization reported in Sec. 7.2.

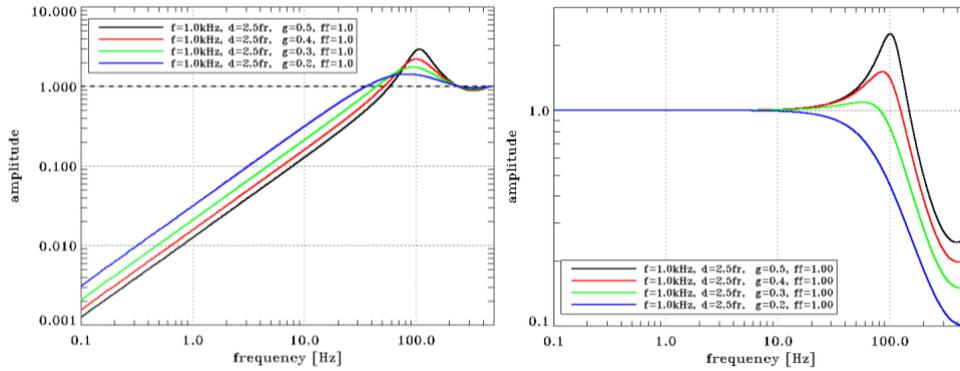

Figure 2. Transfer Functions (TF) of the AO closed loop for a pure integrator control filter. Rejection TFs (RTF) are on left and Noise TFs (NTF) are on right. The plant is considered as a pure delay of 2.5ms.

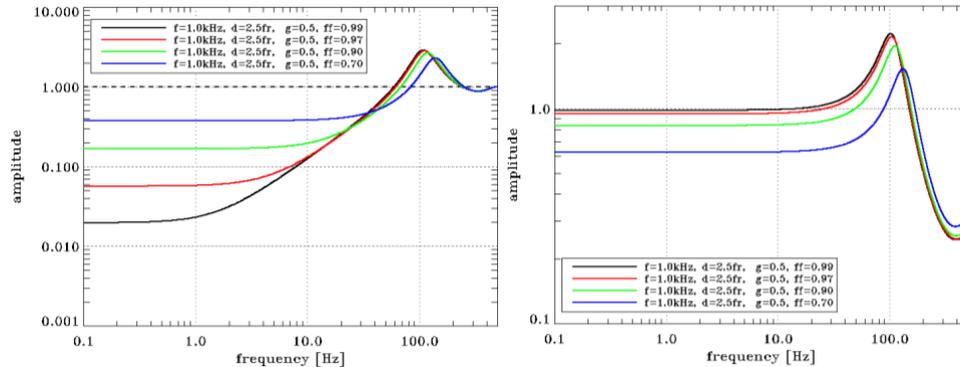

Figure 3. Transfer Functions (TF) of the AO closed loop for a leaky integrator control filter. Rejection TFs (RTF) are on left and Noise TFs (NTF) are on right. The plant is considered as a pure delay of 2.5ms.

# 4. SOUL COMMISSIONING RESULTS

During December 2018 run of the SOUL commissioning [6] we have spent few hours testing the leaky integrator control in the SOUL RTC. This RTC is based on few MVMs and the Adaptive Secondary Adaptive Mirror (ASM) [15] command vector, $c$, is computed as:

$$m(k) = Am(k-1) + GBs(k) \qquad (8)$$
$$c(k) = Mm(k) \qquad (9)$$

where $m$ is the modal command vector, $s$ is the slope vector, $A$ is the state update matrix, $G$ is the gain vector, $B$ is the reconstruction matrix and $M$ is the modes-to-commands matrix. By default, $A$ is an identity matrix, but simply changing the diagonal values leaky integrators can be implemented. We set the diagonal of this matrix as shown in Figure 4. So, we left the first 50 modes with pure integrators and then we decreased the forgetting factor value linearly with radial order. Note that this does not follow the results of the optimization found in Sec. 7.2, because we were not focused on optimization of these coefficients since this was the first test, but instead we wanted a more robust control (see phase margin consideration in Sec. 3) able to face calibration induced errors and different working conditions (seeing, guide star magnitude, wind speed, number of corrected modes, …). We plan to try forgetting factor values optimized as shown in Sec. 7.2 during future tests.

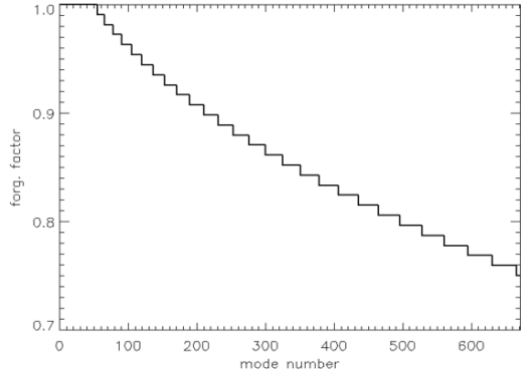

Figure 4. Forgetting factor values used during SOUL commissioning.

The first part of these few hours of commissioning was spent during daytime with a calibration source and the ASM mimicking atmospheric disturbance. We have compared the pure integrator control with the leaky integrator one for relatively faint guide star magnitudes (R=13-14). Table 1 summarizes the results: leaky integrator gives always better results and in particular shows its greater robustness at magnitude R=14.3 when pure integrator is not able to conclude successfully the bootstrap phase of closing the loop (it asked for too large ASM actuator strokes). More results are presented in Figure 5: the modal decomposition of atmospheric disturbance and AO residual on left part of this figure shows that the correction of the leaky integrator is more effective in particular on high order modes and on right part LUCI [16] PSFs in J band show the gain in SR and contrast.

Table 1. SOUL daytime commissioning test results (Pure Integrator, P.I., Leaky Integrator. L.I.).

| control | P.I. | L.I. | P.I. | L.I. | P.I. | L.I. | P.I. | L.I. | P.I. | L.I. | P.I. | L.I. |
|---|---|---|---|---|---|---|---|---|---|---|---|---|
| R | 13 | | 13.5 | | 13.5 | 13.8 | 13.5 | 13.8 | 14.3 | | 14.3 | |
| Seeing [″] | 0.6 | | 0.6 | | 0.6 | | 1.0 | | 0.6 | | 1.0 | |
| No. modes | 300 | 500 | 300 | 500 | 300 | 500 | 300 | 500 | 300 | | 300 | |
| Freq. [Hz] | 500 | | 250 | | 500 | | 500 | | 250 | | 250 | |
| SR (H) | 51.1 | 56.5 | 47.2 | 50.9 | 52.0 | 53.0 | 40.9 | 40.7 | - | 35.5 | - | 30.6 |

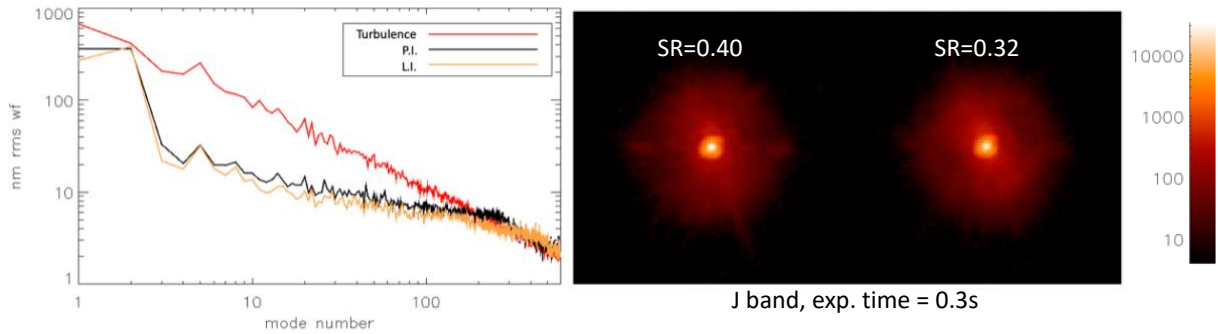

Figure 5. SOUL daytime commissioning test results. Left: modal turbulence and residual (Pure Integrator, P.I., Leaky Integrator, L.I.), right: J band LUCI PSFs. Left PSF is obtained with leaky integrator control, while right with pure integrator control.

Then, we tried the leaky integrator on nighttime. On 18th-19th December 2018 night, we use as guide star BD+20 1790 and during the observation the leaky integrator gives higher performance (see Figure 6): LUCI PSF with FeII filter has a SR of about 70% with an advantage of few percent for the short 🐟 memory control. In particular, the leaky integrator performance corresponds to smaller forces (proportional to inter-actuator stroke and electric current) on the ASM as can be seen in Figure 7. This is very useful to avoid saturation of the ASM forces which could bring to a degradation of the performance and stability issues. Note that this reduction of forces corresponds to a lower power consumption of the voice coil actuators. In particular, referring to the data shown in Figure 7, the power consumption given by the leaky integrator is 45% less than the pure integrator. Our 🐟 is not only gold, but even a bit **green**.

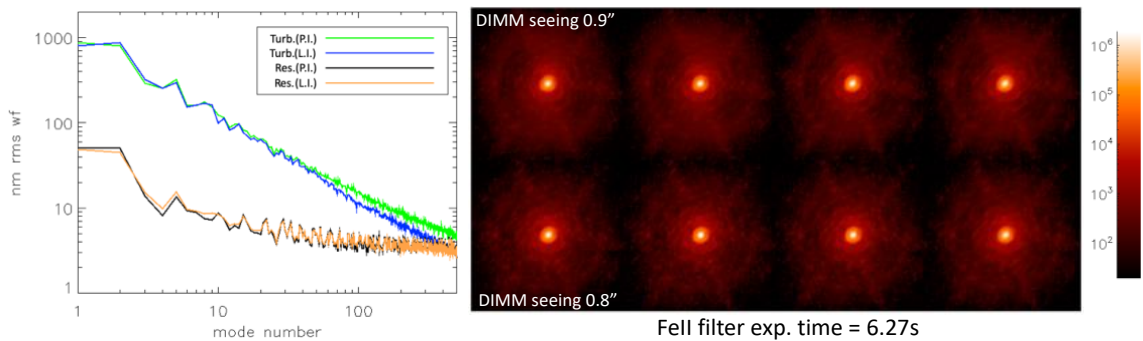

Figure 6. SOUL nighttime commissioning results (guide star BD+20 1790). Left: modal turbulence and residual (Pure Integrator, P.I., Leaky Integrator, L.I.). Note that modal turbulence is estimated from pseudo open loop commands. Right: Fe II filter LUCI PSFs. Top PSFs are obtained with pure integrator control, while bottom PSFs with leaky integrator control.

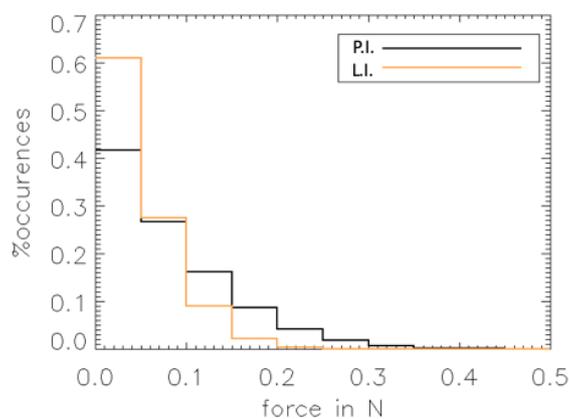

Figure 7. DM force (proportional to inter-actuator stroke and electric current) histogram showing that the leaky integrator (L.I., orange line) occurrence is significantly moved towards lower force values, and, consequently power consumption of voice coil actuators is reduced by 45%.

## 5. FUTURE

During ERIS Adaptive Optics module [17] design phase we have studied the effect of a spatial filter and leaky integrators on the 40×40 sub-aperture SHS of the Natural Guide Star (NGS) mode. We have run end-to-end numerical simulations to evaluate the AO performance with these tools. The results obtained are shown in Figure 8. As expected, the spatial filter is able to reduce aliasing on all modes, but leaky integrators reduce aliasing on high order modes even further if spatial filter is used. In fact, we can see that (on the left part of Figure 8) closed loop residual are lower when the spatial filter is used (green and orange lines) and with leaky integrators they keep decreasing up to the last corrected mode (blue and orange lines) instead of increasing after mode ~500 (red and green lines). Moreover, (on the right part of Figure 8), we see that leaky integrators increase contrast in the PSF between 0.4 and 0.8 arcsec of about a factor 2 (blue and orange lines). So, we found that leaky integrators are complementary to spatial filter, and they do not exclude each other.

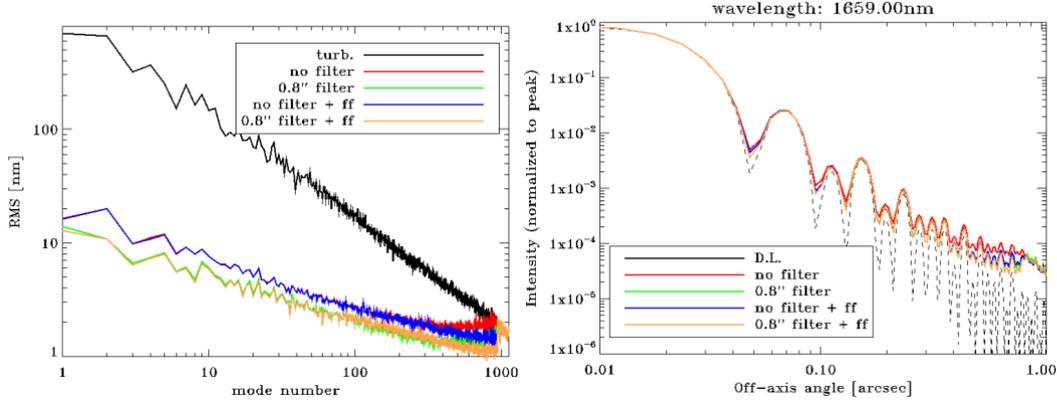

Figure 8. Numerical simulation result for ERIS. Left: open and closed loop modal decomposition with and without SHS spatial filter and with ("ff" in the legend) and without leaky integrators. Right: PSF (@1659nm) profiles for the same cases as left part of this figure.

## 6. CONCLUSION

Elephant or Goldfish? The answer is both! Our control should combine the 🐘 and 🐟 features. In fact, while the pure integrator is still the best choice for low order modes where the turbulence disturbance ratio is high, leaky integrator is preferable for medium/high order modes where the turbulence disturbance ratio is low. The forgetting factor of the leaky integrator is effective in reducing disturbance propagation at low temporal frequencies, improving AO performance and reducing significantly intra-actuator stroke required by the DM without sacrificing the correction quality. To get these advantages, a modal control approach is required, because a zonal control will not benefit from the leaky integrators being dominated by low orders turbulence.

Finally, we showed that not only PWFS, but also SHS (with or without spatial filter) can benefit from it. Actually, the advantage for SHS should be larger than PWFS because it generally suffers more from spatial aliasing (see Sec. 7.1 and [18]).

## 7. APPENDIX

**7.1 Aliasing and Calibration induced errors estimation**

Here we describe the way we used to estimate aliasing and calibration induced errors. We used the end-to-end AO simulation software PASSATA [19]. We set up two different method to estimate aliasing.

In the first method, turbulence is perfectly corrected up to a certain mode (spatial frequency) and the WFS measures this high order phase, so that WFS measurement are only aliasing. In this way we estimate aliasing in different conditions as it is shown in Figure 9 for Pyramid WFS (PWFS) and Figure 10 for Shack-Hartman WFS (SHS). As expected, aliasing is always lower for PWFS, but its variation with seeing ($\varepsilon$) follows a $(\varepsilon/\varepsilon_0)^{5/6}$ law only for SHS, while for PWFS this law is true for modes around no. 400, it is larger for lower order modes and it is smaller for higher order modes (see Figure 11). Note that aliasing PSD has the same features as the turbulence, as can be seen in Figure 12.

In the second method, a WFS without measurement noise corrects the turbulence in closed loop. The actual residual phase is compared with the noiseless WFS measurement: the difference, except for a linear coefficient (the WFS sensitivity), gives an estimation of the aliasing. The WFS sensitivity can be estimated as ratio between the RMS of the noiseless WFS measurement and the RMS of the residual phase. This kind of simulations can be used to estimate the calibration induced error too: for example, introducing a shift between the DM actuators and WFS sub-apertures pattern, we get a larger difference between the noiseless WFS measurement and the actual residual phase as can be seen in the left part of Figure 13. Note that aliasing estimated with this second method is larger by factor in the range 2-4 (see right part of Figure 13). This difference can be caused by the non-perfect AO correction of this second method and by any estimation error of the WFS sensitivity.

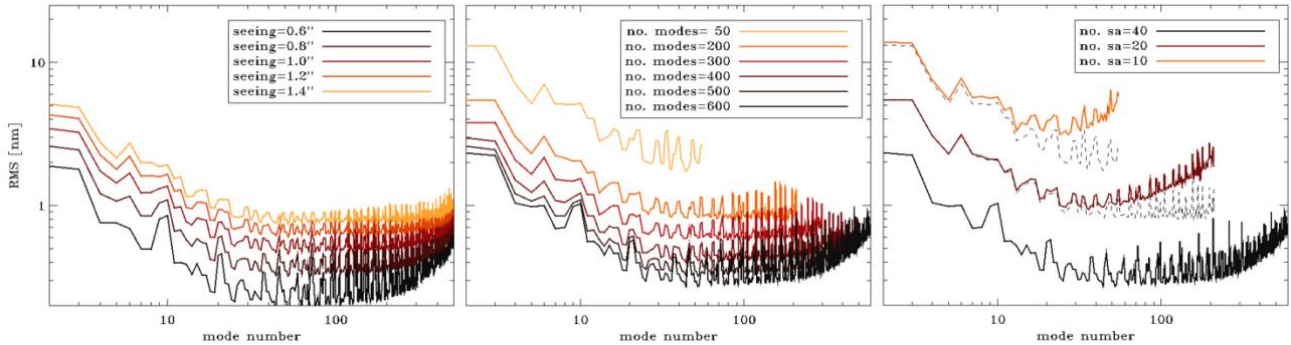

Figure 9. Aliasing RMS for different level of seeing and for a 40x40 sub-aperture PWFS with a modulation of ±3λ/D (computed using first estimation method). Left, different seeing value, center, different number of corrected modes, and, right, binning modes 1, 2 and 4 (which correspond to different pupil samplings). Dashed lines on the right part correspond to aliasing for 40 sub-apertures case with same number of corrected modes (200 and 50) of the lower sampling cases.

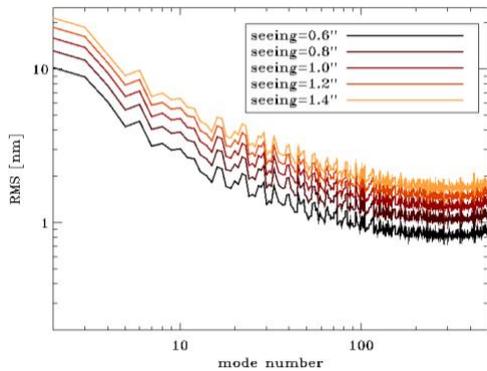

Figure 10. Aliasing RMS for different level of seeing and for a 40x40 sub-aperture SHS (computed using first estimation method). There is about 5 times more aliasing on low order modes and 2 times more aliasing on high order modes than PWFS (correction on 500 modes).

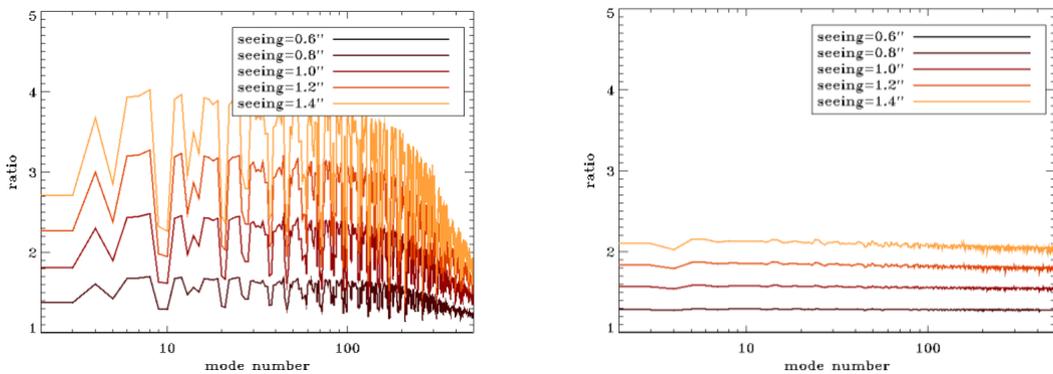

Figure 11. Aliasing RMS for different level of seeing and for a 40x40 sub-aperture PWFS with a modulation of ±3λ/D (left) and SHS (right) (computed using first estimation method).

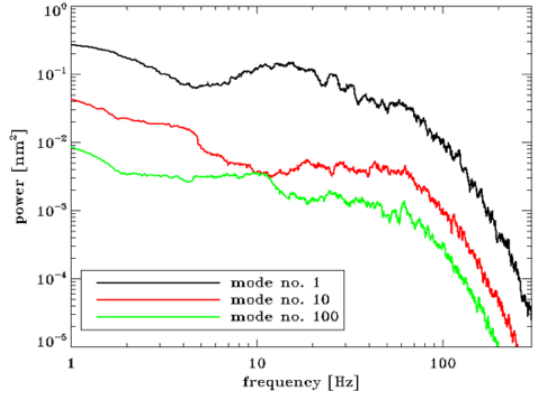

Figure 12. Aliasing PSDs of few modes for seeing 0.8" 500 corrected modes and for a 40x40 sub-aperture PWFS with a modulation of ±3λ/D (computed using first estimation method).

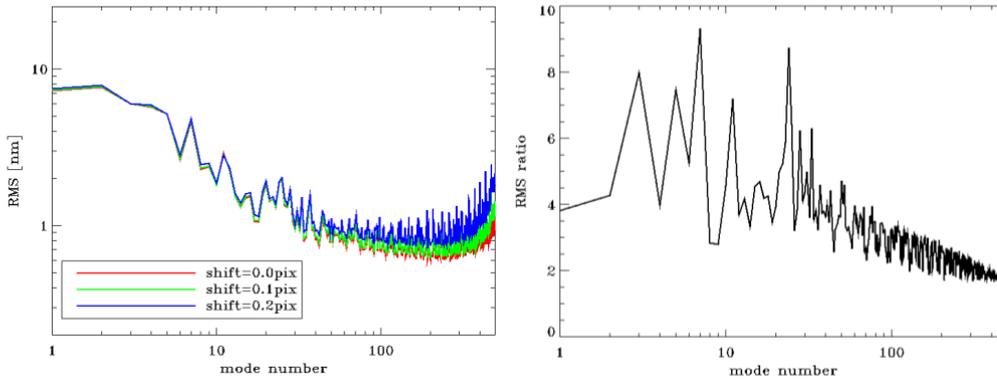

Figure 13. Left: aliasing plus calibration (misalignment on one axis) induced error for a 40x40 sub-aperture PWFS with a modulation of ±3λ/D (computed using second estimation method). Right: RMS ratio between aliasing estimated with second and first method.

**7.2 Forgetting Factors optimization**

In this section we report examples of the optimization of the control filter parameters: gain and forgetting factor. We use the error budget tool described in [7] to make this optimization and the aliasing value estimated with second method shown in Sec. 7.1. Note that we considered no calibration induced errors. The results are reported in Figure 14:

- left part for different seeing values and for a bright star, R=9. We can see that gain is directly proportional to seeing, while forgetting factors decrease when seeing increases. Gain increase to fight stronger turbulence and, so, a greater SNR (flux is the same), but also to oppose the larger loss of sensitivity of the PWFS [20].
- right part for different pupil samplings. Here we chose different configurations to consider realistic cases (taken from [7]): guide star magnitude for 40 sub-apertures case is R=9, for 20 sub-apertures case is R=14.5 and for 10 sub-apertures case is R=16.5. Even frequency of the loop changes decreasing from 1.5kHz of the bright star case to 300Hz of the fainter ones. So, it is not trivial to comment the change of gain values, but it is interesting to see that forgetting factors move faster from 1 when mode number increases for lower pupil sampling. A smaller sampling, and higher star magnitude correspond to a smaller ratio turbulence-aliasing and turbulence-noise requiring a more 🐟 memory to avoid a too large disturbance propagation in the closed loop.

We have fitted the forgetting factor curves in function of the mode number $x$, as it is shown in Figure 15 and reported below:

- Binning mode 1 (40×40 sub-ap.): $y = 2 - e^{(a_1 x)^2}, a_1(\varepsilon = 0.6") = 4.7 \times 10^{-4},\ a_1(\varepsilon = 1.4") = 6.0 \times 10^{-4}$
- Binning mode 2 (20×20 sub-ap.): $y = e^{-a_2 x - b_2 x^2}, a_2 = 6.0 \times 10^{-4}, b_2 = 1.5 \times 10^{-6}$
- Binning mode 4 (10×10 sub-ap.): $y = e^{-a_4 x}, a_4 = 1.5 \times 10^{-3}$

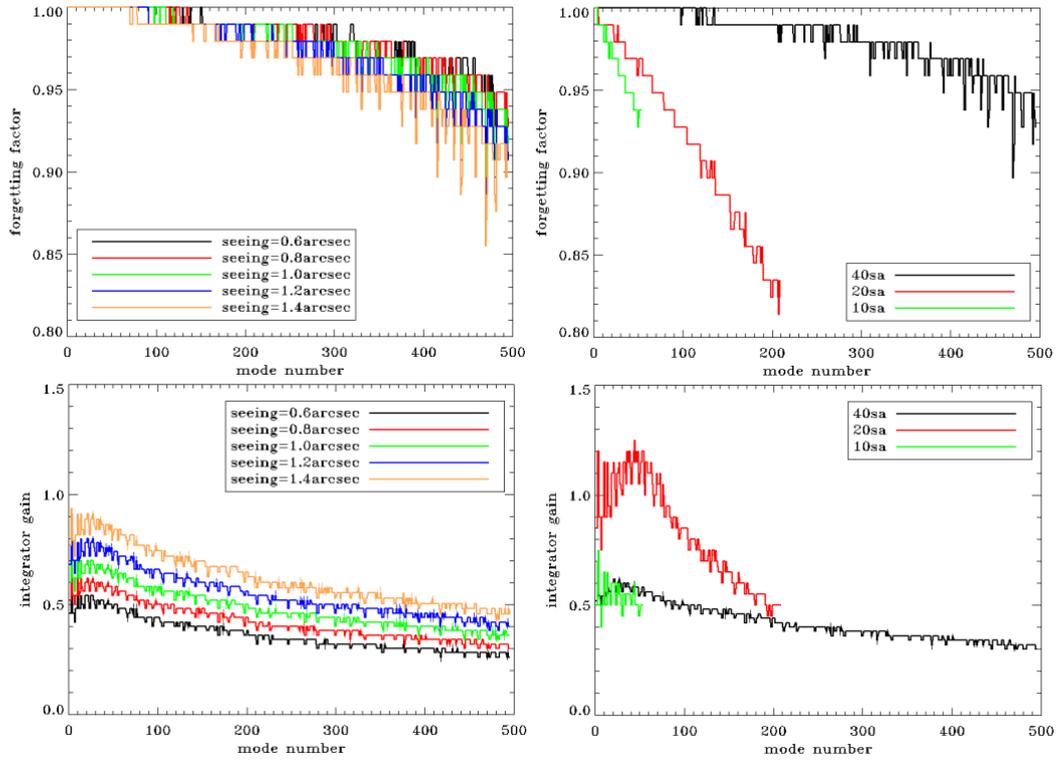

Figure 14. Optimized forgetting factor (top) and integrator gain values (bottom). Left, for different seeing values (0.6÷1.4arcsec) and for a bright star (R=9) and, right, for different number of sub-apertures (40, 20 and 10, that are binning modes 1, 2 and 4). Guide star magnitude considered for 40 sub-apertures case is 9, for 20 sub-apertures case is 14.5 and for 10 sub-apertures case is 16.5.

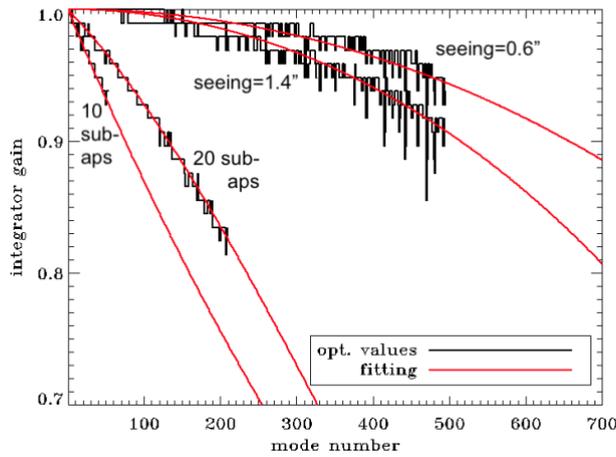

Figure 15. Fitting of the optimized forgetting factor values for different number of sub-apertures (40, 20 and 10, that are binning modes 1, 2 and 4).

## ACKNOWLEDGEMENTS

We wish to acknowledge the help provided by Samuele Calzone in advising citations in the humanities.